\documentstyle[11pt,newpasp,twoside,epsf]{article}
\def\edcomment#1{\iffalse\marginpar{\raggedright\sl#1\/}\else\relax\fi}
\reversemarginpar
 
\begin{document}
\title{High Velocity Clouds and the Local Group}
 \author{Kui-Yun Huang and Ing-Guey Jiang}
\affil{Institute of Astronomy, National Central University, Chung-Li, Taiwan}
 
\begin{abstract}
It was proposed by Blitz et al. (1999) that High Velocity
Clouds (HVCs) are remnants of Local Group formation and 
the average distance of HVCs is 1 Mpc, which is the result of a simple 
dynamical calculation leading to match the observed HVCs distributions.
However, in this paper, we clearly show that fitting the 
observed HVCs distributions by a dynamical calculation {\it cannot} 
provide any constraints on the average distance of HVCs. 
 With our choices of initial conditions, 
the observational results in Wakker \& van Woerden (1991) are produced in our
simulations for the models of both Galactic and 
extragalactic origins.  
Moreover, because Zwaan \& Briggs (2000) reported that 
they failed to locate any extragalactic
counterparts of the Local Group HVCs
in a blind HI 21-cm survey of extragalactic groups,
we  propose to use ``remnants of galactic disc formation''
as the modification for the picture of 
``remnants of galaxy group formation'' in Blitz et al. (1999) and
 thus reduce the possible average
distances of HVCs to be 
about or less than a few hundred kpc.

\end{abstract}
 
\section{Introduction} 
High Velocity Clouds (HVCs) are HI clouds at velocities incompatible
with a simple model of Galactic rotation. 
 
Most HI clouds are the tracers of the galactic discs for
both the Galaxy and the extra-galaxies. 
The reason why HVCs are
interesting is that they do not belong to the majority of the HI clouds
in the Milky Way and therefore their nature and origins need to be
understood.

The well-known proposed HVCs' possible origins usually fall into two categories:
(a) The Local Group Formation -- the remnants of galaxy group formation, 
(b) Galactic Fountain -- the material injected from the galactic
disc "after" the disc was formed. 
HVCs' distances are the most important key parameters to understand 
their possible origins.


Blitz et al.(1999) argued that HVCs should be extragalactic because it is
difficult to understand why HVCs would not be metal rich or how the
vertical
velocities could be so high in the Galactic Fountain context. 
Incidently,
their simulation, which seems to produce the observed
HVCs' distribution on the sky, suggested the mean HVC distance is about 1
Mpc.

On the other hand, Zwann $\&$ Briggs(2000)
reported the result different form
the hypothesis in Blitz et al. (1999).
In a blind HI 21-cm survey of extragalactic group,
they failed to locate any extragalactic counterparts of
the Local Group HVCs.
 
\begin{figure}[t]
\epsfysize 400 pt
\plotone{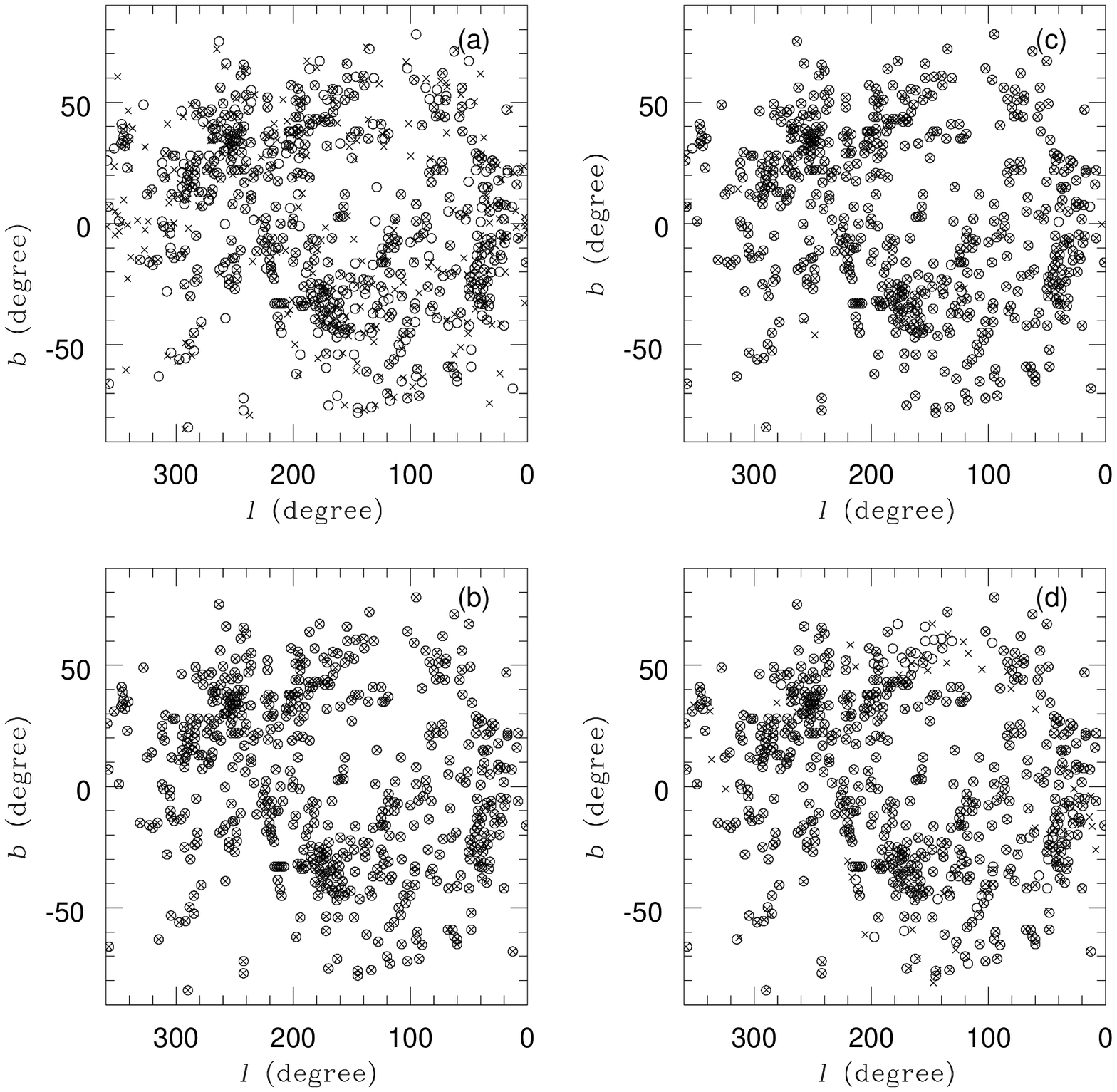}
\caption{The  $l-b$ plots for all models, where circles are observational
data from Wakker \& van Woerden (1991) and cross-points are 
our simulational results:
(a) The Model of Galactic Origins, (b) The Model of Extragalactic Origins,
(c) The Mixed Model, (d) The Model of HVC Complexes.
 }
\end{figure} 

\begin{figure}[t]
\epsfysize 400 pt
\plotone{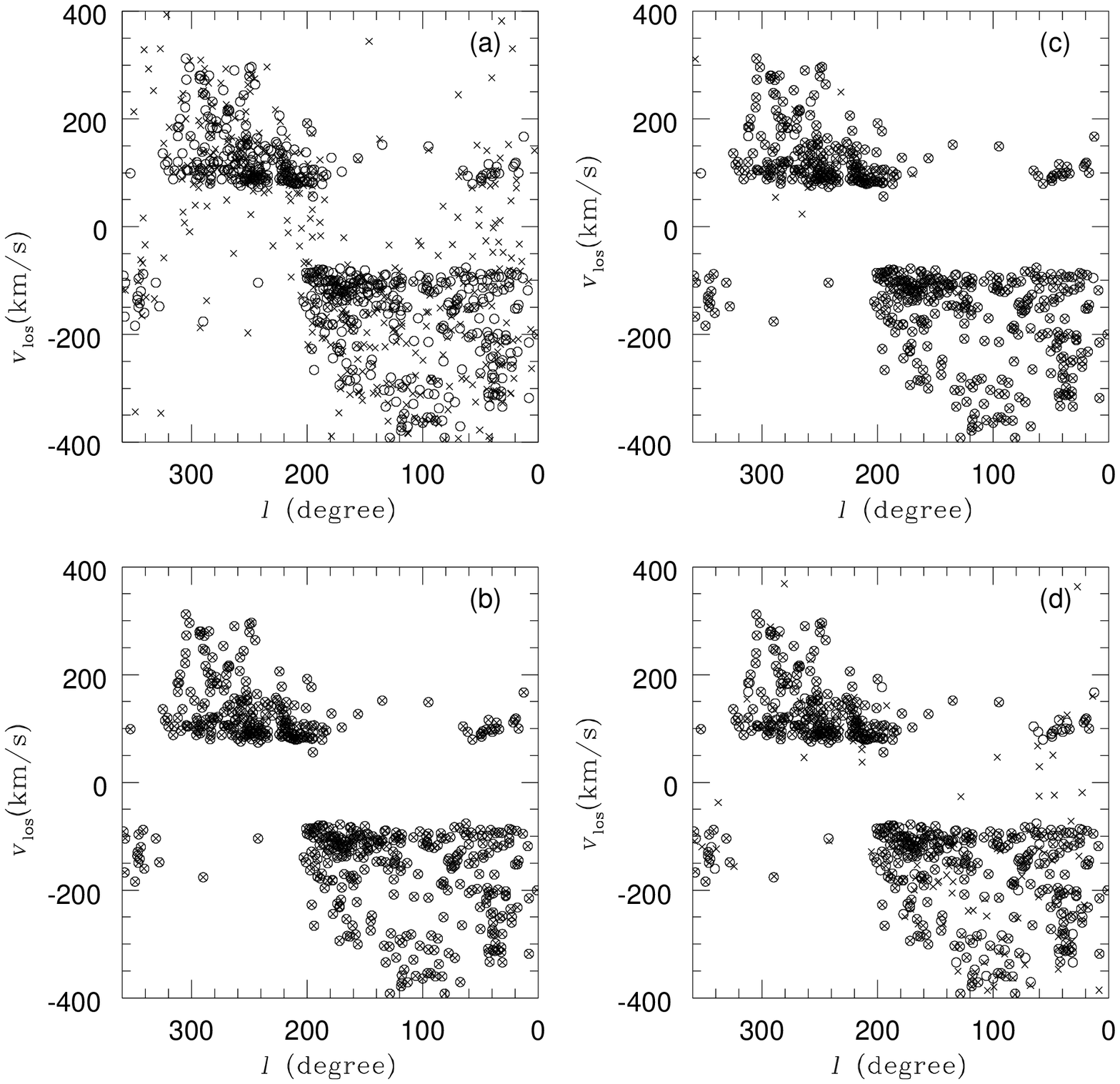}
\caption{The  $l-v_{\rm los}$ 
plots for all models, where circles are observational
data from Wakker \& van Woerden (1991) and cross-points are 
our simulational results:
(a) The Model of Galactic Origins, (b) The Model of Extragalactic Origins,
(c) The Mixed Model, (d) The Model of HVC Complexes.
Please note that those cross-points with 
$|v_{\rm los}| < 50$ km/s should be ignored since they 
do not satisfy the usual definition of HVCs.
}
\end{figure}

\section{The Simulations and Results}

In order to solve this contradiction, we use N-body simulations to model
the HVCs under the gravitational influence from the Milky Way and M31. 
We explore
different initial conditions to check if the simulations can give
constraints on the average distance of HVCs.

Different from what was tried in Blitz et al. (1999), we hope to link 
the observed quantities of each HVC to the simulational results even 
more closely. That is, we wish to reproduce the observed distributions
as perfect as we can. 
In order to have an excellent guess for initial conditions, we plan to 
do the backward integration from the current positions and velocities
of HVCs implied by the observational data.

The observational data in Wakker $\&$ van Woerden (1991) provides us
the present Galactic longitude (${\it l}$), 
Galactic latitude (${\it b}$)
and line-of-sight velocity (${v}_{\rm los}$) of each HVC but there is 
no information about the tangential velocity $v_t$ and Galactic-centre 
distance $R$ since they are more difficult to be determined. 
Therefore, we need to assign both the current values of $v_t$ and $R$ to each
HVC is our simulations.
The tangential velocity $v_t$ is always
determined by a random number between 0 and 20 km/s and its direction 
on the  tangential plane was chosen randomly.

Because the distances to HVCs are the most important quantities to be
understood,
we thus focus on this and plan to explore different range of 
Galactic-centre distance R:
(a) The Model of Galactic Origins (0$< R <$ 50 kpc); 
(b) The Model of Extragalactic Origins (560 $< R <$ 760 kpc); 
(c) The Mixed Model (0$< R <$ 960 kpc);
(d) The Model of HVC Complexes, where many HVCs' $R$  are assumed 
to be within the possible distance-intervals of known  
  HVC Complexes studied in van Woerden et al. (1999),  Wakker (2001), 
Wakker et al. (1998), Ryan et al. (1997), Tamanaha (1996),
and also Wakker \& van Woerden (1991). 
 
We then do the backward integrations to get the initial conditions
for the usual forward integrations. Both were performed for 11 Gyrs.

Figure 1-2 are our simulational results. Figure 1
is the $l-b$ plot and Figure 2 is the $l-v_{\rm los}$ plot.
 
We found that most HVCs arrive the observed positions on the sky
by the end of simulations. We also reproduce the observational HVC distribution
on $l-v_{\rm los}$ plane.
Thus, general features of the observed HVC 
distribution on the sky and $l-v_{\rm los}$ plane
were reproduced  though there are some
deviations. 

\section{Conclusions}

We found that most HVCs' observational distribution on the sky ($l-b$ plane)
and also $l-v_{\rm los}$ plane can be produced for all our different models, 
including both Galactic origin and extragalactic origin models.
Therefore, we conclude that fitting the 
observed HVCs distributions by a dynamical calculation {\it cannot} 
provide any constraints on the average distance of HVCs.

The mechanisms to produce HVCs of Galactic origins can be the galactic fountain
model, stream of satellite galaxies (as the possible stream from Sagittarius 
dwarf galaxy in Figure 6 of Jiang \& Binney 2000) 
and also remnants of galaxy group formation.

On the other hand, the mechanisms to produce HVCs of 
extragalactic origins can be only 
remnants of galaxy group formation unless these HVCs are all close to M31 and 
made by the mechanisms of Galactic origin but within M31.

It is easy to see that there will be no constraint on the distance at all 
if HVCs are the remnants of galaxy group formation and 
thus this suggestion seems
to be able to explain the origins of most HVCs. Especially, if these HVCs
have low-metallicity, other models of Galactic origins would be less 
attractive. Therefore, although one should keep in mind that it is 
always a correct concept that HVCs are
multiple origins, the  ``remnants of galaxy group formation'' is probably 
a good term to describe most HVCs when one has to make a choice.

However,  Zwaan \& Briggs (2000) make a point that there should not be so many
HVCs in the intergalactic space as predicted by Blitz et al. (1999) and give
the picture of ``remnants of galaxy group formation'' a difficult time.


Therefore, in order to resolve the contradiction between 
Blitz et al. (1999) and Zwaan \& Briggs (2000), 
we propose that HVCs are ``remnants of galactic disc formation''
and reduce the possible average
distances of HVCs to be 
about or less than a few hundred kpc.

This conclusion implied from our simulations completely agrees  with 
recent observational results by Tufte et al. (2002), in which
they observed $H_{\alpha}$ lines of HVCs which are
candidates for being at larger than average distance 
and found these clouds are in the
Galactic halo and not distributed throughout the Local Group.
 
\section*{Acknowledgment}
Ing-Guey Jiang wishes to acknowledge the hospitality of Doug Lin during
his visit at UC Santa Cruz in 2000, where he listened Leo Blitz's talk
and began to figure out this project.
This work is supported in part
by the National Science Council, Taiwan, under Grants NSC
90-2112-M-008-052.
 

 
\end{document}